\documentclass{appolb}
\usepackage{epsfig}

\begin{document}
\title{Fragmentation of the fireball\\ 
and how to observe it%
\thanks{Presented at 37th International Symposium on Multiparticle Dynamics, Aug.4-9, 2007, 
Berkeley, California}%
}
\author{Boris Tom\'a\v{s}ik$^{a,b}$, Ivan Melo$^c$, Giorgio Torrieri$^d$,\\ Igor Mishustin$^e$,
Pavol Barto\v{s}$^b$, Mikul\'a\v{s} Gintner$^{b,c}$,\\ Samuel Kor\'ony$^b$ 
\address{$^a$ FNSPE, Czech Technical University, Prague, Czech Republic\\
$^b$ Univerzita Mateja Bela, Bansk\'a Bystrica, Slovakia\\
$^c$ \v{Z}ilinsk\'a Univerzita, \v{Z}ilina, Slovakia\\
$^d$ Institut f\"ur theoretische Physik, Universit\"at Frankfurt,\\ Frankfurt am Main, Germany\\
$^e$ Frankfurt Institute for Advanced Studies, Frankfurt am Main, Germany}
}
\maketitle
\begin{abstract}
We argue that fragmentation at hadronisation is likely scenario
in ultrarelativistic nuclear collisions. In case of crossover phase transition it
is driven by a singularity of the bulk viscosity. We claim that such a scenario can 
explain the ``HBT puzzle'' and can be identified by non-statistical differences 
between event-wise rapidity distributions and by proton-proton rapidity correlations.
\end{abstract}
 
 
\textbf{Introduction.}
In ultrarelativistic heavy ion collisions we aim at creating deconfined and chirally restored matter.
Even if that goal is reached, the system expands dramatically and 
eventually undergoes transition to hadronic phase. From lattice QCD
we know that at vanishing and/or small baryochemical potential the transition is a rapid though 
smooth crossover \cite{aoki}. The crossover becomes sharper as the baryochemical potential 
increases and turns into a first order phase transition at a critical point. 

It is rather well known that if a system expands very fast through a first order phase 
transition it supercools and fragments via spinodal decomposition. Fragmentation, however, 
can also occur in case of rapid crossover \cite{oil2honey}. 
The culprit for this is singular behaviour of bulk viscosity near $T_c$ 
\cite{Kharzeev:2007wb,Karsch:2007jc}. 

Fragmentation would affect measured sizes of the fireball, event-wise rapidity 
spectra, and clustering would be seen in rapidity correlations \cite{pratt94,randrup}


\textbf{Fragmentation in rapid phase transitions.}
First order phase transitions run via nucleation initiated on impurities or
thermal fluctuations. In general, time is needed to nucleate critical bubbles by 
thermal fluctuation. In rapidly expanding systems, if the expansion rate is bigger 
than the rate for critical bubble nucleation large supercooling can occur \cite{supercool}. 
The system can reach the spinodal point in which mechanical instability leads to 
fragmentation \cite{spinodal}.

It may seem that fragmentation scenario is irrelevant for heavy ion collisions at RHIC
and LHC which run in the region of phase diagram where smooth 
crossover appears. However, even in this regime bulk viscosity as a function of 
temperature shows singular behaviour at $T_c$ \cite{Kharzeev:2007wb,Karsch:2007jc}. It is 
negligible otherwise. Bulk viscosity appears in the term $\zeta \partial_\mu u^\mu$ and
thus scales the reluctance of the fluid to change its volume; large $\zeta$ means 
that the system resists against fast changes of the volume. 

Hence, from the beginning of the collision the fireball expansion accelerates and large 
expansion velocity is built up. Then, at $T_c$ suddenly large bulk viscosity appears and makes the 
fireball ``rigid'', i.e.\ not willing to expand. On the other hand,  inertia  
tries to keep the expansion going. As a result of these two competing effects the bulk may 
break up into fragments if its inner forces cannot hold it together anymore. In \cite{oil2honey}
typical size of fragments was estimated for Bjorken one-dimensional expansion from 
energy considerations
\begin{equation}
\label{dsize}
L^2=\frac{24\zeta_c\tau_c}{\varepsilon_c}\, ,
\end{equation}
where $\tau_c$ and $\varepsilon_c$ are the proper time and energy density at $T_c$,
and $\zeta_c$ parametrizes the singular behaviour of bulk viscosity 
$\zeta(\tau) = \zeta_c \delta(\tau - \tau_c)$.

After the fragmentation, final-state hadrons evaporate from fragments.


\textbf{Observable consequences: femtoscopy.} 
A failure of hydrodynamic simulations to reproduce the measured correlation radii is 
known as the ``HBT puzzle'' \cite{pratt}. Simulations yield the outward 
correlation radius $R_o$ much too big in comparison with the sideward radius $R_s$. In terms 
of second-order spatial moments of the source 
$
R_o^2 = \langle \tilde x^2 \rangle  - 2\frac{K_t}{K_0}\langle \tilde x \tilde t\rangle +
\frac{K_t^2}{K_0^2} \langle \tilde t^2 \rangle \, , 
$
where $K$ is the average pair momentum and the $x$-coordinate is directed parallel to $K_t$
(tilde denotes coordinates w.r.t. mean position of the source). The correlation radii are 
determined by the size, orientation, and shape of the freeze-out hypersurface. A typical 
freeze-out hypersurface from hydrodynamic simulation leads to negative $\langle \tilde x \tilde t\rangle$
term and thus increases $R_o$. In a scenario with freeze-out from fragments, hadrons are emitted 
from a different interval of the space-time and this could solve the ``HBT puzzle'' \cite{wong,oil2honey}.

\begin{figure}[tb]
\centerline{\epsfig{file=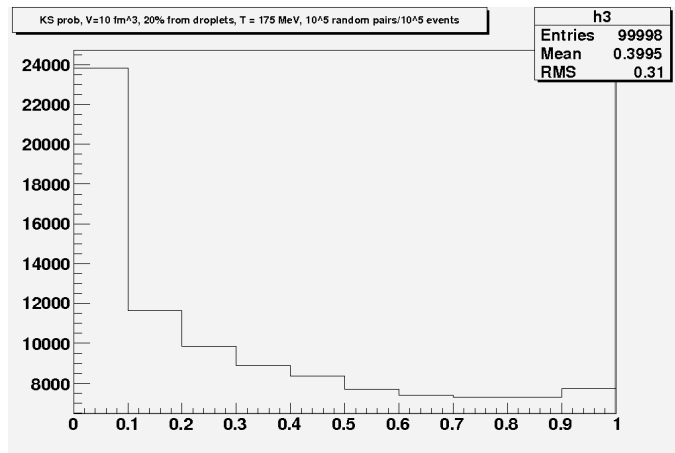,width=9.0cm}}
\caption{\label{f-KS} 
Typical result of Kolmogorov-Smirnov test on a sample of 10$^5$ pairs out of 10$^5$ events
events in which 20\% of hadrons are emitted from fragments with average volume 10~fm$^3$. 
}
\end{figure}

\textbf{Event generator for droplet emission.}
In order to investigate various observables which could be measured in case of 
fireball fragmentation a Monte Carlo generator has been developed which 
generates positions and momenta of hadrons. In its spirit it is similar to 
THERMINATOR \cite{therminator}, though partices are emitted from fragments. This leads to 
clustering in momentum space, since particles emitted from one fragment inherit 
their velocities close to that of the fragment. For the results presented here
no resonance decays were included. 


\textbf{Observable consequences: event-wise rapidity distributions.}
If the fireball disintegrates then emitted particles will have rapidities close
to those of the fragments. Therefore, there will be (possibly overlapping) clusters in 
hadronic rapidity distributions. Rapidities of the fragments will differ from event to 
event. Thus each event will be given by to different rapidity distribution. On the other hand,
if there is no fragmentation then in a sample of carefully centrality-selected events
rapidity distributions in each event will be the same. 

In statistical sense, we can ask to what extent two sets of measured rapidities 
from two events look like coming from the same underlying distribution. A standard tool 
for such a study is Kolmogorov-Smirnov test. An example of our use of the test is in Figure~\ref{f-KS}. 
For shortness we can only mention that flat distribution would correspond to 
all events looking alike, while a peak at 0 indicates non-statistical differences between events. 
The signal is very clear. This study will be reported in a forthcoming paper. 


\begin{figure}[tb]
\centerline{\epsfig{file=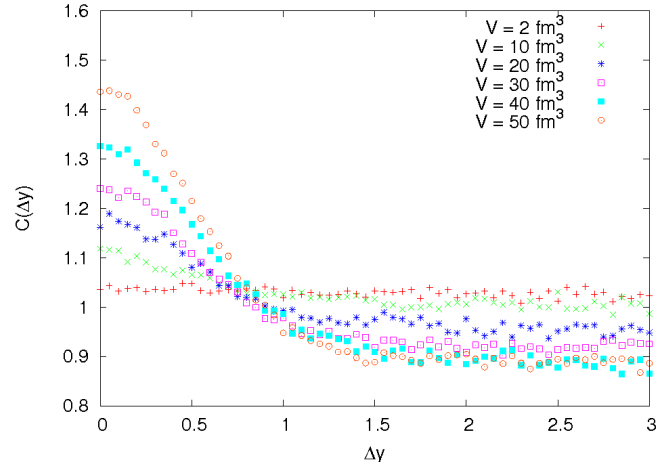,width=9.0cm}}
\caption{\label{f-pcor} 
Proton-proton correlation function in rapidity for varying average fragment sizes. All hadrons are
emitted from droplets. Fermi-Dirac statistics and $pp$ interaction have not been taken into account. 
Correlation functions are not normalised.}
\end{figure}

\textbf{Observable consequences: rapidity correlations.}
It has been also suggested that droplets should lead to a contribution to proton-proton correlation function
in rapidity \cite{pratt94,randrup}. Such correlation functions are shown in Figure \ref{f-pcor}. 
We clearly observe  that the visibility of the signal increases with the size of droplets
(note that total multiplicity was kept constant in these simulations).

This work was supported by  MSM 6840770039 and LC 07048
(Czech Republic), VEGA 1/4012/07 (Slovakia), 
DFG grant 436RUS 113/711/0-2 (Germany), as well as
RFFR-05-02-04013 and NS-8756.2006.2 (Russia).


\end{document}